\documentclass[a4paper, 12pt]{article}

\usepackage[english]{babel}
\usepackage[utf8x]{inputenc}
\usepackage[T1]{fontenc}

\usepackage[paper=letterpaper,margin=.85in]{geometry}

\usepackage{amsmath}
\usepackage{graphicx} %allows you to use jpg or png images. PDF is still recommended
\usepackage[colorlinks=true,linktocpage]{hyperref} % add links inside PDF files
\usepackage{cleveref}
\usepackage{xcolor}
\usepackage{amsfonts} 
\usepackage{amssymb}  
\usepackage{authblk}
\usepackage{cite}
\usepackage{tensor}
\usepackage[export]{adjustbox}
\usepackage{url}

\hypersetup{linkcolor = blue, citecolor=red}

\DeclareMathOperator{\tr}{tr}
\DeclareMathOperator{\re}{Re}
\DeclareMathOperator{\im}{Im}

\makeatletter
\newcommand\appendix@section[1]{%
  \refstepcounter{section}%
  \orig@section*{Appendix \@Alph\c@section: #1}%
  \addcontentsline{toc}{section}{Appendix \@Alph\c@section: #1}%
}
\let\orig@section\section
\g@addto@macro\appendix{\let\section\appendix@section}
\makeatother

\newcommand{\beq}{\begin{equation}}
\newcommand{\eeq}{\end{equation}}
\newcommand{\p}{\partial}

\newcommand{\nn}{\nonumber}
\newcommand{\la}{\langle}
\newcommand{\ra}{\rangle}

\newcommand{\bJ}{\bar J}

\title{\bf Factorization and complex couplings \\ in SYK and in Matrix Models }

\author{ Baur Mukhametzhanov}
\affil{\it Institute for Advanced Study \\ \it Princeton, NJ 08540, U.S.A.}

  \date{}

\begin{document}
\maketitle
\thispagestyle{empty}

\vskip 1in

\begin{abstract}

We consider the factorization problem in toy models of holography, in SYK and in Matrix Models. In a theory with fixed couplings, we introduce a fictitious ensemble averaging by inserting a projector onto fixed couplings. We compute the squared partition function and find that at large $N$ for a typical choice of the fixed couplings it can be approximated by two terms: a ``wormhole'' plus a ``pair of linked half-wormholes''. This resolves the factorization problem. 

We find that the second, half-wormhole, term can be thought of as averaging over the imaginary part of the couplings. In SYK, this reproduces known results from a different perspective. In a matrix model with an arbitrary potential, we propose the form of the ``pair of linked half-wormholes'' contribution. In GUE, we check that errors are indeed small for a typical choice of the hamiltonian. Our computation relies on a result by Brezin and Zee for a correlator of resolvents in a ``deterministic plus random'' ensemble of matrices.

\end{abstract}

\clearpage
\pagenumbering{arabic} 
\tableofcontents
\clearpage

\section{Introduction}

In this work we study the factorization problem \cite{Maldacena:2004rf} in toy models of AdS/CFT \cite{Maldacena:1997re, Gubser:1998bc, Witten:1998qj}. We would like to compute a two-boundary observable, e.g. a product of two partition functions $z_L(J) z_R(J)$, corresponding to the ``left'' $L$ and ``right'' $R$ boundaries. In the bulk computation, there are contributions from wormhole spacetimes connecting the two boundaries. Such contributions seem to spoil factorization between $L$ and $R$. This is not a problem in a theory averaged over the couplings $J$, which are the same on all boundaries. But it requires an explanation in a theory with fixed couplings. 

\vskip .3in

This question is particularly important because wormholes play a significant role in explaining the long-time physics of various quantities of interest: the spectral formfactor in two \cite{Saad:2018bqo, Saad:2019lba} and in higher \cite{Cotler:2020ugk, Cotler:2020lxj, Cotler:2021cqa} dimensions, correlation functions \cite{Saad:2019pqd, Blommaert:2020seb}, entropy of Hawking radiation \cite{Penington:2019kki, Almheiri:2019qdq}, squared matrix elements \cite{Stanford:2020wkf}.

\vskip .3in

Recently, the factorization problem was studied in the SYK model \cite{Sachdev_1993, Kitaev1, Kitaev2, Maldacena:2016hyu, Kitaev:2017awl} by Saad, Shenker, Stanford and Yao \cite{Saad:2021rcu}. They computed the squared partition function in a theory with fixed couplings. They showed that at large $N$ there are two important contributions. First, there is a wormhole contribution that is approximately the same as in the averaged theory. Second, there is a contribution they called a ``pair of linked half-wormholes''. Together, these two terms are a good approximation in the theory with fixed couplings and factorization is restored
\begin{align}\label{zzIntro}
z_L(j)z_R(j) \approx \la z_L(J) z_R(J) \ra_J + \Phi(j) \ .
\end{align}
This equation is written schematically and will be made precise in the main text. In the LHS $j$ denotes the fixed couplings of the theory, e.g. $j_{a_1 \dots a_q} $ in the SYK model without averaging. In the first term in the RHS $J$ denotes the averaged couplings and $\la \dots \ra_J$ is the averaging over $J$. This term gives the wormhole contribution. It does not depend on the fixed couplings $j$. The second term $\Phi(j)$, linked half-wormholes, depends on the couplings erratically and averages to zero $\la \Phi(j) \ra_j = 0$. Similar phenomena was observed \cite{Saad:2021uzi} in the topological gravity model of Marolf and Maxfield \cite{Marolf:2020xie}.

\vskip .3in

To derive \eqref{zzIntro}, the authors in \cite{Saad:2021rcu} noticed that one can introduce $G,\Sigma$ collective fields in the SYK model with fixed couplings. The fields $G, \Sigma$ depend on both $L$ and $R$ degrees of freedom and are a proxy for the bulk variables. Using these collective fields, \cite{Saad:2021rcu} showed that there are two saddle points corresponding to the two terms in \eqref{zzIntro}.

\vskip .3in

A surprising feature of \eqref{zzIntro} is that the second term, the pair of linked half-wormholes, has a reasonably simple description in the semi-classical limit. 

\vskip .3in

It is interesting to ask how general is the approximation \eqref{zzIntro}. In partiucular, what is the analog of \eqref{zzIntro} in other theories with ensemble averages, such as matrix models. It is not immediately clear how to generalize the analysis of \cite{Saad:2021rcu}, because there is no obvious analog of the $G, \Sigma$ description.

\vskip .3in

We are particularly motivated to understand \eqref{zzIntro} in matrix models because their gravity duals have been extensively studied, recently \cite{Saad:2019lba, Stanford:2019vob, Witten:2020wvy, Maxfield:2020ale, Mertens:2020hbs, Turiaci:2020fjj, Blommaert:2019wfy, Okuyama:2019xbv, Okuyama:2020ncd, Okuyama:2021eju, Suzuki:2021zbe} and over the last few decades \cite{David:1984tx, Ambjorn:1985az, Kazakov:1985ds, Kazakov:1985ea, Brezin:1990rb, Douglas:1989ve, Gross:1989vs, Seiberg:2003nm, Maldacena:2004sn}. See \cite{Ginsparg:1993is, DiFrancesco:1993cyw, Seiberg:2004at, Nakayama:2004vk} for review.

\subsection{Summary of results}
\label{Summary}

In this work we present an alternative way of deriving the approximation \eqref{zzIntro}, that doesn't rely on the existence of the collective fields $G, \Sigma$. In particular, our method works equally well in both SYK and matrix models. 

\vskip .3in

The main idea is to consider the theory with fixed couplings $j$ as an ensemble averaged theory with an insertion of a delta function of the couplings $\delta(J-j)$. The delta function can in turn be written as an integral over the second set of couplings $\sigma$
\begin{align}
z(j)^2 &= \int dJ ~ \delta(j-J) e^{V(j) - V(J) } z(J)^2  \\
&= \int {dJ d\sigma \over 2\pi} e^{ V(j) - V(J) + i \sigma(J-j) }  z(J)^2 \\
&= e^{V(j) }\int {d\sigma \over 2\pi} ~ e^{- i \sigma j} \la z(J)^2 e^{i\sigma J} \ra_J \ ,
\label{zzDeltaIntro}
\end{align}
where the potential $V(J)$ corresponds to the choice of the ensemble\footnote{Here, we illustrate the idea and are not careful about specifying the theory or the number of different couplings. The details in specific examples will be worked out in subsequent sections.}. In the last line we introduced the ensemble averaging $\la \cdot \ra_J = \int dJ ~ e^{-V(J)} (\cdot)$.

\vskip .3in

In order to obtain \eqref{zzIntro}, we will find that we can approximate in \eqref{zzDeltaIntro}
\begin{align}\label{zzWIntro}
{ \la z(J)^2 e^{i\sigma J} \ra \over \la e^{i \sigma J} \ra } \approx \la z(J)^2 \ra + { \la z(J) e^{i\sigma J} \ra \over \la e^{i\sigma J}\ra} ~ { \la z(J) e^{i\sigma J} \ra \over \la e^{i\sigma J}\ra} \ ,
\end{align}
where we divided by factors of $\la e^{i\sigma J} \ra$ to normalize the averages in the presence of external couplings $\sigma$. Our motivation for the second term is that it has a similar pattern of correlations as linked half-wormholes in SYK \cite{Saad:2021rcu, Mukhametzhanov:2021nea} and in the Marolf-Maxfield model \cite{Saad:2021uzi}. There are no direct $J$ correlations between two $z$'s, but they are ``linked'' through correlations with $e^{i\sigma J}$.

\vskip .3in

As we will show in the particular examples, the equation \eqref{zzWIntro} is valid in the limit of large Hilbert space dimension $L$ (or large number of Majorana fermions $N$ in the SYK model). We can also continue to long Lorentzian time where the wormhole contribution gives rise to the linear growth in time (the ramp) \cite{Cotler:2016fpe}. The equation \eqref{zzWIntro} is still valid in that regime and resolves the factorization puzzle. The first term gives the average linear growth, while the second term gives large oscillations around the average. 

One can imagine going to even longer times that are exponential in the entropy $t \sim e^S$, the so-called plateau regime \cite{Cotler:2016fpe}. In this case it is known that the wormhole receives important non-perturbative corrections \cite{Cotler:2016fpe} that stop the linear in time growth. Similarly, we expect that the second term in the RHS of \eqref{zzWIntro} must be corrected in order to resolve the factorization puzzle in the plateau region.

\vskip .3in

In section \ref{SYK} we describe the above computation in detail in SYK and SYK with one time point and recover the linked half-wormholes from \cite{Saad:2021rcu, Mukhametzhanov:2021nea}. In particular, we find that the half-wormhole term can be thought of as averaging over the imaginary part of the couplings. For example, in the SYK model with one time point
\begin{align}
z(j)^2 \approx \la z^2 \ra + \la z(j + i \delta \sigma)^2 \ra_{\delta \sigma} \ .
\end{align}
The imaginary part of the couplings $\delta \sigma$ is the shifted $\sigma$ variable from \eqref{zzDeltaIntro}: $\sigma = j + i \delta \sigma$. A similar expression in the full SYK is derived in \cref{SYKfull}.

\vskip .3in

In section \ref{MatrixModels} we consider Matrix Models. Going through steps similar to \eqref{zzDeltaIntro} and \eqref{zzWIntro}, we propose an approximation for the squared partition function \eqref{HWapproxMM} for a general potential $V(H)$. In the gaussian case $V(H) = {1\over 2}H^2$, we show that it is indeed a good approximation. Here, our computation crucially relies on the two-point correlator of resolvents, derived by Brezin, Hikami and Zee \cite{Brezin:1993lnq,Brezin:1994sq,Brezin:1995fs}, in a ``deterministic plus random'' ensemble of matrices. We leave checking our proposal for a more general form of the potential $V(H)$ for future work, but it seems reasonable to expect that it holds in this case as well.

\vskip .3in

We end with a discussion and open questions in section \ref{discuss}.

\vskip .3in

Similar ideas have been recently explored in \cite{Blommaert:2021gha}. There, instead of sharply fixing the couplings with a delta function, the authors smear it a little bit and study the dependence of matrix models on the width of the smearing. The complex couplings in SYK model have been also discussed in \cite{Garcia-Garcia:2021squ, Garcia-Garcia:2020ttf}.

\section{SYK}
\label{SYK}

\subsection{SYK with one time point}
\label{SYK1t}

In this section we study the SYK model with one time point \cite{Saad:2021rcu, Mukhametzhanov:2021nea}
\begin{align}\label{sykdef}
z(J)= \int d^N\psi ~ \exp\left( i^{q/2} J_A \psi_A \right) \ .
\end{align}
We use capital latin letters $A,B,C, \dots$ to denote ordered $q$-subsets of $\{ 1,\dots , N\}$. For example
\begin{align}
A = \{ a_1 < \dots < a_q \} \ , \qquad J_A \psi_A \equiv J_{a_1 \dots a_q} \psi_{a_1 \dots a_q}  \equiv J_{a_1 \dots a_q}  \psi_{a_1} \dots \psi_{a_q} \ ,
\end{align}
where $\psi_a$ are Grassmann numbers. The completely antisymmetric tensor of the couplings $J_A = J_{a_1 \dots a_q}$ is drawn out of a gaussian ensemble with zero mean and variance 
\begin{align}\label{Jdist}
\la J_A J_B  \ra = \bJ^2 \delta_{AB} \ , \qquad 
\bJ^2 \equiv {(q-1)! \over N^{q-1}} \ .
\end{align}
We assume $N,q$ are both even integers.

\vskip .3in

We would like to study the squared partition function with fixed couplings $j_A$. In order to isolate the wormhole contribution and study the factorization problem, we introduce a fictitious averaging over couplings 
\begin{align}\label{InsertDelta1}
z(j)^2 &= \int \prod_A \Big[ dJ_A ~ \delta(J_A - j_A) \Big] ~ \exp \left\{-{1\over 2\bar J^2} (J_A^2 - j_A^2) \right\} ~ z(J)^2 \\
& = \int \prod_A \left( dJ_A d\sigma_A \over 2\pi \bar J^2 \right) ~ 
\exp \left\{-{1\over 2\bar J^2} (J_A^2 - j_A^2) + {1\over \bar J^2 } i \sigma_A(J_A - j_A) \right\} ~ z(J)^2 \ . 
\label{InsertDelta2}
\end{align}
In the second line, we wrote delta-functions as integrals over $\sigma_A$ with a normalization that will be convenient below. Let us denote the normalized SYK averaging as
\begin{align}\label{sykav}
\la \cdot \ra_J = \int \prod_A \left( dJ_A \over \sqrt{2\pi} \bar J \right) e^{-{1\over 2\bar J^2} J_A^2} (\cdot) \ .
\end{align}
Whenever clear, we leave the averaged couplings implicit and simply write $\la \cdot \ra$. Now the equation \eqref{InsertDelta2} takes the form
\begin{align}\label{factorizedSFF}
z(j)^2 &=  \int \prod_A \left( d\sigma_A \over \sqrt{2\pi} \bar J \right) ~ 
\exp\left\{ - {(\sigma_A + i j_A)^2 \over 2 \bar J^2} \right\} ~ { \left\la z(J)^2 W( J) \right\ra \over \la W(J) \ra } \ , \\
W(J) &\equiv \exp\left\{ i \sigma_A J_A\over \bar J^2 \right\}  \ .
\end{align}
Here, we multiplied and divided by a factor of $\la W \ra$, which we will find to be convenient below, and used that 
\begin{align}\label{Wav}
\la W \ra = \exp \left\{ - {\sigma_A^2 \over 2 \bar J^2 } \right\} \ .
\end{align}
From the bulk perspective, the insertion of $W(J)$ in the average can be thought of as creating baby universes \cite{Marolf:2020xie, Saad:2021uzi} that can be connected to the boundaries created by $z(J)^2$.

\vskip .3in 

The formula \eqref{factorizedSFF} is, of course, a trivial rewriting of the theory with fixed couplings $j_A$. But now we would like to ask: is there an approximation of $\la z^2 W \ra$, such that \eqref{factorizedSFF} holds to leading order at large $N$? We will find that the following turns out to be a good approximation
\begin{align}\label{z2Wapprox}
{ \la z^2 W \ra \over \la W \ra } \approx \la z^2 \ra +  { \la z W \ra \over \la W \ra } { \la z W \ra \over \la W \ra } \ .
\end{align}
The first term in the RHS is the ``wormhole'' contribution corresponding to the disconnected part $\la z^2 W \ra \supset \la z^2 \ra \la W \ra $. In the second term in the RHS we consider contributions where each $z$ is correlated with $W$, but there are no correlations between two $z$'s.

\vskip .3in

The reader might find it puzzling at first glance that there are two factors of $\la W \ra$ in the denominator in the second term of \eqref{z2Wapprox}. This can be thought of as a normalization factor in the ensemble deformed by $W$. And the second term is the disconnected contribution in this deformed ensemble. In more detail, we can understand it as follows. When we compute Wick contractions with $W$, the final result contains a factor of $\la W \ra$. For example, $\la J_A W \ra = \la J_A  {i \sigma_B J_B \over \bar J^2 } \ra \la W \ra$. In the LHS of \eqref{z2Wapprox} we divided by $\la W \ra$ to remove this factor. In the second term in the RHS we have two correlators and need to remove two factors of $\la W \ra$.

\vskip .3in

The non-trivial content of the approximation \eqref{z2Wapprox} is that in general two $z$'s could be correlated with each other and with $W$ at the same time. We neglected such terms in \eqref{z2Wapprox}.

\vskip .3in

Our motivation for including the second term in \eqref{z2Wapprox} is that it has a similar pattern of correlations as the ``pair of linked half-wormholes'' considered in \cite{Saad:2021rcu, Mukhametzhanov:2021nea, Saad:2021uzi}. There are no direct $J_A$ correlations between two $z$'s, but they are ``linked'' through correlations with $W$.

\vskip .3in

Similarly to \cite{Saad:2021rcu, Mukhametzhanov:2021nea}, we will find that the approximation \eqref{z2Wapprox} is valid for a {\it typical} choice of the couplings $j_A$. In other words, the variance of $z(j)^2$ around the approximation \eqref{z2Wapprox} is suppressed at large $N$, if $j_A$'s are drawn from the same ensemble as $J_A$'s. We will discuss this in more detail below. In fact, the second term in \eqref{z2Wapprox} will precisely reproduce the half-wormhole contribution from \cite{Saad:2021rcu,Mukhametzhanov:2021nea}. 

\vskip .3in

Using the definition \eqref{sykdef} we compute
\begin{align}
\la z W \ra & = \int d^N\psi \left\la \exp\left\{ i^{q/2} J_A \psi_A +  {i\sigma_A J_A \over  \bar J^2 } \right\} \right\ra \\
& = \int d^N\psi  \exp\left\{ i^{q/2}~ i \sigma_A \psi_A - {\sigma_A^2\over  2\bar J^2 }  \right\} \\
& = z(i\sigma) \la W \ra \ .
\label{zW}
\end{align}
Inserting \eqref{z2Wapprox}, \eqref{zW} into \eqref{factorizedSFF} we find
\begin{align}\label{z2approxIm}
z(j)^2 \approx \la z^2 \ra +   \int \prod_A \left( d\sigma_A \over \sqrt{2\pi} \bar J \right) ~ 
\exp\left\{ - {(\sigma_A + i j_A)^2 \over 2 \bar J^2} \right\} ~z(i\sigma)^2 \ . 
\end{align}
It is curious that the integration over $\sigma$ is with the same measure as the SYK averaging over couplings \eqref{sykav}, but with an overall shift by $j_A$.\footnote{This is, of course, mainly due to the fact that we consider a gaussian ensemble \eqref{sykav}, for which the measure is invariant under the fourier transformation \eqref{Wav}. It is not entirely clear how much of this would survive if we add non-gaussianities or when we consider matrix models with arbitrary potentials. In SYK and in GUE, we will find formulas analogous to \eqref{z2approxIm}, \eqref{ImJ}. One might expect that the variables $\sigma_A$ might play an interesting role along similar lines in more general cases. } Defining $\delta \sigma_A$ by $i \sigma_A = j_A + i \delta \sigma_A $ we can write
\begin{align}\label{ImJ}
z(j)^2 \approx \la z^2 \ra + \la z(j + i \delta \sigma)^2 \ra_{\delta \sigma} \ ,
\end{align}
where $\la \cdot \ra_{\delta \sigma}$ denotes the averaging over ``couplings'' $\delta \sigma$ with the measure identical to the standard SYK averaging \eqref{sykav}. Thus, the half-wormhole contribution (we will show why it is a half-wormhole momentarily) can be thought of as averaging over the imaginary part $\delta \sigma_A$ of the couplings $j_A + i\delta \sigma_A$, while the real part $j_A$ corresponds to the fixed couplings in the LHS of \eqref{ImJ}.

\vskip .3in

Using \eqref{sykdef}, a straightforward computation gives 
\begin{align}
 \la z(j + i \delta \sigma)^2 \ra_{\delta \sigma} = \int d^{2N} \psi ~ \exp\left\{ i^{q/2} j_A (\psi_A^L + \psi_A^R) - i^q \bar J^2 \psi_A^L \psi_A^R \right\} \ .
\end{align}
This is precisely the half-wormhole contribution considered in \cite{Saad:2021rcu,Mukhametzhanov:2021nea}, where it was obtained using the $G\Sigma$ effective action. As discussed there, the approximation \eqref{ImJ} holds for typical couplings $j_A$ in the sense that 
\begin{align}\label{ErrorSYK1pt}
\la \text{Error}(j) \ra_j = 0 \ , \qquad 
{ \sqrt{ \la |\text{Error(j)}|^2  \ra_j } \over \la z^2 \ra } \ll 1 \ ,
\end{align}
where Error is defined as the difference of the LHS and RHS of \eqref{ImJ}
\begin{align}
\text{Error}(j) \equiv z(j)^2 - \la z^2 \ra -  \la z(j + i \delta \sigma)^2 \ra_{\delta \sigma} \ .
\end{align}
We interpret \eqref{ErrorSYK1pt} as follows. The first equation means that the approximation \eqref{ImJ} holds for a typical, bit fixed, choice of the couplings $j$. The second equation computes the variance and means that corrections to \eqref{ImJ} are small for typical couplings $j$. The equations \eqref{ErrorSYK1pt} were derived in \cite{Saad:2021rcu,Mukhametzhanov:2021nea}.

\subsection{SYK}
\label{SYKfull}

We now turn to the one-dimensional SYK model \cite{Sachdev_1993, Kitaev1, Kitaev2, Maldacena:2016hyu, Kitaev:2017awl}. The only new feature is that we have to deal with the average of the partition function itself $\la Z(\beta) \ra$ (disc contribution in the putative bulk dual), that was absent in the previous section. We can either subtract it by hand or go to the long lorentzian times, where it decays to zero. 

\vskip .3in

The partition function is defined as\footnote{Here we emphasize the dependence on the couplings $j_A$ by writing $Z(j)$ and omit the dependence on $\beta$. Sometimes we will write $Z_L(j)$ or $Z_R(j)$, implying that it depends on $\beta_L$ or $\beta_R$ respectively for the ``left'' and ``right'' systems. }
\begin{align}\label{SYKdef}
Z(J) = \int D \psi ~ \exp \left\{ - \int_0^\beta d\tau \left( \psi_i \p_\tau \psi_i + i^{q/2} J_A \psi_A \right) \right\} \ ,
\end{align}
where $\psi_a(\tau)$ are Majorana fermions and the couplings $J_A$ are drawn from the same gaussian distribution \eqref{Jdist}. We again have the representation \eqref{factorizedSFF} of the spectral formfactor with the fixed couplings $j_A$
\begin{align}\label{FactorizedSFF}
Z_L(j) Z_R(j) &=  \int \prod_A \left( d\sigma_A \over \sqrt{2\pi} \bar J \right) ~ 
\exp\left\{ - {(\sigma_A + i j_A)^2 \over 2 \bar J^2} \right\} ~ { \left\la Z_L(J) Z_R(J)  W(J) \right\ra \over \la W(J) \ra } 
\end{align}
and make the approximation analogous to \eqref{z2Wapprox}
\begin{align}\label{Z2Wapprox}
{ \la  Z_L Z_R W \ra \over \la W \ra } \approx \la  Z_L Z_R \ra_c +  { \la Z_L W \ra \over \la W \ra } { \la Z_R  W \ra \over \la W \ra } \ .
\end{align}
Here, the first term in the RHS is the connected part of the correlator, i.e. the wormhole contribution. The completely disconnected part $\la Z_L \ra \la  Z_R \ra$ is contained in the second term. 

\vskip .3in

Using \eqref{SYKdef} we compute
\begin{align}
{ \la Z  W \ra \over \la W \ra } &= 
 \int D \psi ~ \exp \left\{ - \int_0^\beta d\tau \left( \psi_i \p_\tau \psi_i + i^{q/2} i \sigma_A \psi_A \right) + 
 {\bar J^2 \over 2 } i^q  \iint d\tau d\tau' ~\psi_A(\tau) \psi_A(\tau') \right\}  \\
 & = \la Z(J+ i \sigma) \ra_J \ .
 \label{ZW}
\end{align}
In the second line we noted that the $\psi$ path integral in the first line can be thought of as the partition function with the imaginary part of the couplings $i \sigma_A$ fixed and the real part $J_A$ averaged over. This is analogous to \eqref{zW}.\footnote{In SYK with one time point $ \la z(J + i \sigma) \ra_J = z(i\sigma)$. }

\vskip .3in

Combining \eqref{FactorizedSFF},\eqref{Z2Wapprox},\eqref{ZW} and introducing $\delta \sigma_A$ as before $i\sigma_A = j_A + i \delta \sigma_A$, we have
\begin{align}\label{HWapprox}
Z_L(j) Z_R(j)  &\approx \la  Z_L Z_R \ra_c +  \Phi(j) \ , \\
\Phi(j) &= 
 \left\la 
Z_L(j+J^L+i\delta\sigma) 
Z_R(j+J^R+i\delta\sigma)
 \right\ra_{J^L,J^R,\delta \sigma} \ .
\end{align}
The linked half-wormholes contribution\footnote{In the way we defined it, $\Phi(j)$ contains the disconnected piece $\la Z_L \ra \la Z_R \ra$ and would be more appropriately called ``disconnected plus linked half-wormholes''. We will subtract disconnected terms below.} $\Phi(j)$ can also be computed more explicitly as a $\psi$ path integral
\begin{align}
\Phi(j) = \int D\psi^{L,R} ~ 
\exp \Big\{ 
&-\int_0^{\beta_L} d\tau~ (\psi_i^L \p_\tau \psi_i^L  +i^{q/2} j_A \psi_A^L)
-\int_0^{\beta_R} d\tau~  (\psi_i^R \p_\tau \psi_i^R  +i^{q/2} j_A \psi_A^R)\\
&- {\bar J^2 \over 2} i^q \iint d\tau d\tau' ~ \psi_A^L(\tau) \psi_A^R(\tau') 
 \Big\} \ .
\end{align}
Using this path integral, it is straightforward to show that the average of $\Phi$ gives the disconnected contribution
\begin{align}\label{Phiav}
\la \Phi(j) \ra_j = \la Z_L \ra \la Z_R \ra \ .
\end{align}

It is convenient to subtract terms proportional to $\la Z_L \ra, \la Z_R \ra$ from both sides of \eqref{HWapprox}. We define
\begin{align}
z_a(j) = Z_a(j) - \la Z_a \ra , \qquad a = (L,R) \ .
\end{align}
The equation \eqref{HWapprox} can be equivalently written as
\begin{align}\label{HWapprox2}
z_L(j) z_R(j)  &\approx \la  z_L z_R \ra +  \phi(j) \ , \\
\phi(j) &= 
 \left\la 
\left[( Z_L(j+J^L+i\delta\sigma)  - Z_L(J^L) \right]
\left[( Z_R(j+J^R+i\delta\sigma)  - Z_L(J^R) \right]
 \right\ra_{J^L,J^R,\delta \sigma} \ ,
 \label{phiSYKim}
\end{align}
where we used that $Z_a(j)$ can be written in a form similar to \eqref{FactorizedSFF}. Note that the half-wormhole contribution $\phi(j)$ is real if $\beta_L, \beta_R$ are real (or complex conjugates of each other), since the averaging over $\delta \sigma$ is symmetric under $\delta \sigma_A \to - \delta \sigma_A$.

\vskip .3in

Again, the pair of linked half-wormholes \eqref{phiSYKim} can be thought of as the squared partition function averaged over the imaginary part of the couplings $\delta \sigma_A$, up to $\la Z_{L,R} \ra$ subtractions.

\vskip .3in

In order to discuss whether \eqref{HWapprox2} is a good approximation for a typical choice of couplings $j_A$, we consider as before
\begin{align}
\text{Error}(j) &= z_L(j) z_R(j)  - \la  z_L z_R\ra -  \phi(j) \ .
\end{align}
The average of Error vanishes $\la \text{Error}(j) \ra_j = \la \phi(j) \ra_j = 0$. The average of $\text{Error}^2$ is 
\begin{align}\label{Error2SYK}
\la \text{Error}(j)^2 \ra_j =  \la  (z_L z_R)^2 \ra -  \la  z_L z_R \ra^2
-  2 \la    \phi  z_L z_R \ra + \la  \phi^2  \ra  \ .
\end{align}
The first two terms can be computed by the wormhole saddles\footnote{More precisely, in order for the wormhole to be genuine saddle, one has to consider the microcanonical partition function and continue to the lorentzian times. See \cite{Saad:2018bqo} for details.} \cite{Saad:2018bqo} and give to leading order at large $N$
\begin{align}
\la  (z_L z_R)^2 \ra -  \la  z_L z_R \ra^2 \approx \la z_L z_R \ra^2 + \la z_L z_L \ra_c \la z_R z_R \ra \ .
\end{align}
We expect that a similar leading order estimate holds for the other two terms in \eqref{Error2SYK}
\begin{align}\label{dP2}
\la  \phi^2 \ra \approx  \la    \phi z_L z_R \ra \approx 
\la z_L z_R \ra^2 + \la z_L z_L \ra \la z_R z_R \ra \ . 
\end{align}
This equation is motivated by the fact that a similar result was found in the SYK with one time point \cite{Saad:2021rcu,Mukhametzhanov:2021nea}. We will also show it holds in GUE in the next section. If these estimates hold, then the RHS of \eqref{Error2SYK} cancels out to leading order at large $N$ and $\text{Error}(j)$ is indeed small on average
\begin{align}
{\la \text{Error}^2 \ra  \over \la (z_L z_R)^2 \ra } \ll 1 \ .
\end{align}

\vskip .3in

To compute $\la  \phi^2 \ra$ and $\la    \phi z_L z_R \ra $, we can write $\psi$ path integrals and do the averaging. We would then find, similarly to \cite{Saad:2018bqo}, that there are wormhole saddles corresponding to the RHS of \eqref{dP2}. It is, however, difficult to justify that there are no other significant contributions at large $N$\cite{Saad:2021rcu}.

\section{Matrix Models}
\label{MatrixModels}

Now we consider similar phenomena in matrix models. We study a product of two partition functions of a fixed $L \times L$ hermitian hamiltonian $H_0$. Similarly to the previous section, we introduce a fictitious averaging over the ensemble as
\begin{align}
Z_L(H_0) Z_R (H_0)&= \int dH ~  \delta(H-H_0) ~e^{ - L \tr  \left[ V(H) - V(H_0) \right] } ~Z_L(H) Z_R (H) \\
& = \int dH d\Sigma ~e^{ i L \tr \Sigma (H-H_0) - L \tr  \left[ V(H) - V(H_0) \right] } ~Z_L(H) Z_R (H) \ ,
\label{ZZdeltaMM}
\end{align}
where $\Sigma$ is another $L\times L$ hermitian matrix and 
\begin{align}
Z_{a}(H) = \tr e^{-\beta_{a} H} \ , \qquad a = (L,R) \ .
\end{align}
Introducing the averaging
\begin{align}\label{avdef}
\la \cdot \ra_H = \int dH ~ e^{-L \tr V(H)} (\cdot) \ ,
\end{align}
we can write\footnote{Whenever clear, we omit the subscript and denote the averaging by $\la \cdot \ra$.}
\begin{align} \label{Z2int}
Z_L(H_0) Z_R (H_0)&=  e^{ L V(H_0) } \int  d\Sigma ~e^{ - i L \tr \Sigma H_0  } ~ \Big\la Z_L(H) Z_R (H) W(H) \Big\ra \ , \\
W(H) &\equiv e^{ i  L \tr \Sigma H } \ .
\end{align}
So far this is a trivial rewriting of $Z_L Z_R$. Now we can make an approximation analogous to \eqref{z2Wapprox} and \eqref{Z2Wapprox}
\begin{align}\label{ZZW}
{\la Z_L Z_R W \ra \over \la W \ra} \approx \la Z_L Z_R \ra_c + { \la Z_L W \ra \over \la W \ra } { \la Z_R W \ra \over \la W \ra } \ .
\end{align}
Inserting this into \eqref{Z2int}, we find
\begin{align}\label{HWapproxMM}
Z_L(H_0) Z_R (H_0) &\approx
\la Z_L Z_R \ra_c + \Phi(H_0) \ , \\
\Phi(H_0) &= e^{ L \tr V(H_0) }  \int d\Sigma~e^{ - i L \tr \Sigma H_0  } ~ \la W \ra ~ { \la Z_L W \ra \over \la W \ra } { \la Z_R W \ra \over \la W \ra } \ .
\end{align}

In order to slightly simplify the discussion, it is convenient to subtract the ``disk'' contributions $\la Z_L \ra, \la Z_R \ra$ by defining\footnote{Alternatively, following \cite{Saad:2018bqo}, one could consider a microcanonical partition function $Y_{E, \Delta E}(T)$ continued to lorentzian time $T$. In this case the disk contribution decays exponentially in time and can be discarded.  }
\begin{align}
z_a(H) = Z_a(H) - \la Z_a \ra \ , \qquad a = (L,R) \ .
\end{align}
Then the equation \eqref{HWapproxMM} can be equivalently written as
\begin{align}\label{HWapproxMM2}
z_L(H_0) z_R (H_0) &\approx
\la z_L z_R \ra + \phi(H_0) \ , \\
\phi(H_0) &= e^{ L \tr V(H_0) }  \int d\Sigma~e^{ - i L \tr \Sigma H_0  } ~ \la W \ra ~ { \la z_L W \ra \over \la W \ra } { \la z_R W \ra \over \la W \ra } \ .
\label{phiDef}
\end{align}
This is one of our main results: a proposal for the contribution of ``linked half-wormholes'' in a matrix model with an arbitrary potential $V(H)$. Similarly to the SYK model in the previous section, we expect \eqref{HWapproxMM2} to hold for a typical choice of $H_0$ drawn from the ensemble defined by the potential $V(H_0)$. 

\vskip .3in

Note that $\phi(H_0)$ is real if $\beta_L, \beta_R$ are real (in fact $ \beta_L^* = \beta_R$ is enough). Taking complex conjugate in \eqref{phiDef} corresponds to a change of the integration variable $\Sigma \to - \Sigma$.

\vskip .3in

We define Error as the difference between the LHS and the RHS of \eqref{HWapproxMM2} 
\begin{align}\label{ErrorDef}
\text{Error}(H_0) = z_L(H_0) z_R (H_0) -
\left( \la z_L z_R \ra + \phi(H_0) \right) \ .
\end{align}
We expect the approximation \eqref{HWapproxMM2} is valid in the sense that
\begin{align}\label{ErrorAv}
\la \text{Error}(H_0) \ra_{H_0} = 0 \ , \qquad 
\la \text{Error}(H_0)^2 \ra_{H_0} \ll \left\la \left( z_L(H_0) z_R(H_0) \right)^2  \right\ra_{H_0} \ .
\end{align}
Below, we will do detailed computations in the gaussian case $V(H) = {1\over 2} H^2$. We expect that \eqref{ErrorAv} is true for a general form of the potential $V(H)$, though we leave showing this for future work. See the discussion in section \ref{discuss} for more on this.

\vskip .3in

The first equation in \eqref{ErrorAv} requires $\la \phi(H_0) \ra_{H_0} = 0$. This is straightforward from the definition \eqref{phiDef}. We have
\begin{align}\label{phiAv}
\la \phi(H_0) \ra_{H_0}  = \int dH_0 d\Sigma ~e^{ - i L \tr \Sigma H_0  } ~ \la W \ra ~ { \la z_L W \ra \over \la W \ra } { \la z_R W \ra \over \la W \ra } = 0 \ .
\end{align}
Doing first the integral over $H_0$ gives a delta function, which sets $\Sigma = 0$ and $W=1$. This leaves us with $\la z_L\ra \la z_R \ra = 0$.

\vskip .3in

Computing $\la \text{Error}^2 \ra$ is less trivial. From the definition \eqref{ErrorDef} we find
\begin{align}\label{ErrorAv2}
\la \text{Error}^2 \ra = \la (z_L z_R)^2 \ra - \la z_L z_R \ra^2 -  2\la \phi z_L z_R \ra + \la \phi^2 \ra \ .
\end{align}
Here, it is implied that $z_L, z_R, \phi$ are functions of $H_0$ and the averaging is over $H_0$. We would like to show that the four terms in the RHS of \eqref{ErrorAv2} are cancelled to leading order at large $L$. The first term at large $L$ is dominated by three disconnected contributions, one of which cancels the second term, and we can approximate
\begin{align}\label{ErrorAv22}
\la \text{Error}^2 \ra \approx \la z_L z_R \ra^2 + \la z_L^2 \ra \la z_R^2 \ra - 2 \la \phi  z_L z_R \ra + \la \phi^2 \ra \ .
\end{align}
A more interesting part of the computation is to estimate $\la \phi^* \phi \ra$ and $\la \phi z_L z_R \ra$ at large $L$.

\vskip .3in

Below, we will restrict ourselves to the gaussian ensemble $V(H) = {1\over 2} H^2$. We will compute $\la \phi^2 \ra , \la \phi z_L z_R \ra$ and find that\footnote{One might expect the estimate \eqref{phi2Res} is true for an arbitrary matrix potential $V(H)$.}
\begin{align}\label{phi2Res}
  \la \phi^2 \ra  \approx \la \phi z_L z_R \ra \approx \la z_L z_R \ra^2 + \la z_L^2 \ra \la z_R^2 \ra 
\end{align}
to leading order at large $L$, thus showing that the RHS of \eqref{ErrorAv22} cancels out and the typical size of Error is small
\begin{align}\label{ErrorSmall}
{\la  \text{Error}^2 \ra  \over \la (z_L z_R)^2 \ra } \ll 1 \ .
\end{align}

By the standard genus counting \cite{tHooft:1973alw, Brezin:1977sv}, the connected correlator $\la z_L z_R \ra$ is of $O(1)$ at large $L$. We are therefore interested in similar terms in $\la \phi^2 \ra , \la \phi z_L z_R \ra$.

\vskip .3in 

The approximations \eqref{phi2Res} and \eqref{ErrorSmall} are valid at large $L$ for any euclidean times $\beta_L, \beta_R$. Further, we can analytically continue to $\beta_L +it, \beta_R - it$ and consider the regime of long lorentzian time $t$. In particular, in the regime where the wormhole contribution gives the linear growth in time (the ramp) \cite{Cotler:2016fpe}, the above equations resolve the factorization puzzle. The wormhole provides the average, while the half-wormhole $\phi$ is responsible for large oscillations around the average. If we go to even longer times, the so-called plateau regime \cite{Cotler:2016fpe}, we expect non-perturbative corrections to be important both for the averaged wormhole contribution $\la z_L z_R \ra$ as well as the half-wormhole $\phi$.

\vskip .3in

Before discussing the squared partition function \eqref{HWapproxMM2}, it is useful to look at a quantity for which it is easier to check that Error is small. The simplest observable with a factorization problem is $\tr H_0 \tr H_0$ in GUE. In this case, however, the approximation analogous to \eqref{ZZW} is exact and there isn't anything to check. It becomes more interesting if we consider $\tr H_0^2 \tr H_0^2$. In this case the approximation \eqref{ZZW} is non-trivial. In appendix \ref{trH2} we show that an analog of \eqref{HWapproxMM2} for $\tr H_0^2 \tr H_0^2$ is a good approximation.

\subsection{GUE}

We now turn to the special case $V(H) = {1\over 2}H^2$ and compute $\la \text{Error}^2 \ra$ in detail. 

\vskip .3in

First, the average of $W$ is easy to compute in GUE
\begin{align} \label{WavGUE}
\la W \ra = \la e^{iL \tr \Sigma H} \ra = \exp \left\{ - {L\over 2} \tr \Sigma^2 \right\} \ .
\end{align}
This allows us to write the linked half-wormholes \eqref{phiDef} as
\begin{align}
\phi(H_0) &=  \int d\Sigma~e^{ -  {L\over 2} \tr (\Sigma + i H_0)^2  } ~ { \la z_L W \ra \over \la W \ra } { \la z_R W \ra \over \la W \ra } \ .
\label{phiGUE}
\end{align}
It is interesting to note that \eqref{phiGUE} can be written in a form similar to \eqref{phiSYKim}. First, in the gaussian case using \eqref{WavGUE} we have
\begin{align}
{\la ZW \ra \over \la W\ra } &= \int dH ~ e^{ -{L\over 2} \tr (H -i \Sigma)^2 } ~ \tr e^{-\beta H} \\
& = \la Z(H + i\Sigma) \ra_H \ .
\end{align}
This is analogous to \eqref{zW}, \eqref{ZW}. Then \eqref{phiGUE} can be written similarly to \eqref{phiSYKim}
\begin{align}\label{phiGUEim}
\phi(H_0) = 
 \big\la 
\left[ ( Z_L(H_0+H_L+i\delta\Sigma)  - Z_L(H_L) \right]
\left[ ( Z_R(H_0+H_R+i\delta\Sigma)  - Z_L(H_R) \right]
 \big\ra_{H^L,H^R,\delta \Sigma} \ ,
\end{align}
where $H_L, H_R, \delta \Sigma$ are all drawn from a gaussian ensemble with zero mean. We defined $\delta \Sigma$ by $i \Sigma = H_0+ i \delta \Sigma$. Again, the contribution of linked half-wormholes corresponds to averaging over the imaginary part of the hamiltonian $i \delta \Sigma$.

\vskip .3in

Now we compute \eqref{phiGUE}, \eqref{phiGUEim} more explicitly. The correlators of the type $\la Z W \ra$ in the presence of an external field $W = e^{i L \tr \Sigma H}$ have been studied in the literature in the gaussian ensemble \cite{Kazakov:1990nd, Brezin:1996hze, Brezin:1997zz, Brezin:2007aa}, in more general random matrix potentials \cite{Gross:1991aj, Zinn-Justin:1997hrj, 1998CMaPh.194..631Z} and, more recently, in \cite{Blommaert:2021gha} in the context of the factorization problem. The integrand is no longer invariant under unitary transformations. However, using Harish-Chandra--Itzykson--Zuber integral over unitaries \cite{Harish-Chandra, Itzykson:1979fi}, one can reduce the matrix model to an integral over eigenvalues. The latter can be explicitly computed in the gaussian case. The exact result at finite $L$ is (see section 2 in both \cite{Brezin:1996hze} and \cite{Blommaert:2021gha}, which we adapt to our conventions)
\begin{align}\label{ZW GUE}
 { \la Z W \ra \over \la W \ra }  = - {L \over \beta} e^{\beta^2/2L} \oint {du \over 2\pi i } ~e^{-\beta u } \det \left( 1 - {\beta \over L} {1\over u - i \Sigma} \right) \ , 
\end{align}
where the contour encloses all the eigenvalues of $i\Sigma$ counterclockwise. We will be interested in the leading terms at large $L$. Introducing the resolvent $R_M(u)$ of a matrix $M$ by
\begin{align}
R_M(u) = {1\over L} \tr {1\over u - M}
\end{align}
we expand \eqref{ZW GUE} and get\footnote{We normalized the potential for $\Sigma$ such that its eigenvalue density is supported on the interval of $O(1)$ and the resolvent is also $O(1)$ at large $L$. }
\begin{align}
 { \la Z W \ra \over \la W \ra }  &= -{L\over \beta} \oint {du \over 2\pi i } ~e^{-\beta [u + R_{i\Sigma}(u)] } 
 \left(
 1 +
 {\beta^2 \over 2L} (1 + R'_{i \Sigma}(u))
 +  O\left(  L^{-2}  \right) 
 \right)
  \\
  & = -{L\over \beta} \oint {du \over 2\pi i } ~e^{-\beta [u + R_{i\Sigma}(u)] } 
 \left(
 1 
 +  O\left(  L^{-2}  \right) 
 \right) \ . 
 \label{ZWlargeL}
\end{align}
In the second line we dropped the $O(1/L)$ term because it is a total derivative. In particular, the expansion is in powers of $1\over L^2$. 

\vskip .3in

It is instructive to check that for $\Sigma = 0$ the resolvent is $R_{i \Sigma=0}(u) = {1\over u}$ and we recover the Wigner semicircle from \eqref{ZWlargeL}
\begin{align}\label{Sigma=0}
\la Z \ra \approx
-{L\over \beta} \oint {du \over 2\pi i } ~e^{-\beta \left(u + {1\over u} \right)} = L~ {I_1(2\beta) \over \beta} = \int_{-2}^2 dE ~ e^{-\beta E} ~ {L\over 2\pi} \sqrt{4- E^2} \ .
\end{align}
Inserting \eqref{ZWlargeL} into \eqref{phiGUE} we find
\begin{align}\label{phiGUE2}
\phi(H_0) &= {L^2 \over \beta_L \beta_R} \oint {du_L \over 2\pi i} {du_R \over 2\pi i } ~ e^{-\beta_L u_L - \beta_R u_R} 
\nn
\\ 
&\times 
\left\la 
\left( e^{-\beta_L R_{H_0+i\delta\Sigma}(u_L)}  - e^{-\beta_L/u_L} \right)
\left( e^{  -\beta_R  R_{H_0 + i\delta\Sigma}(u_R)  }  - e^{-\beta_R/u_R} \right)
\right\ra_{\delta \Sigma} + \dots \ , 
\end{align}
where we made a shift $i\Sigma = H_0 + i \delta \Sigma$, such that $\delta \Sigma$ averaging is \eqref{avdef} with the gaussian potential $V(\delta \Sigma) = {1\over 2}\delta \Sigma^2$. We also used that the disk partition function $\la Z\ra$ can be written as the $u$-integral \eqref{Sigma=0} and made corresponding subtractions in the integrand in \eqref{phiGUE2}.

\subsection{Ginibre ensemble of complex matrices}
\label{GinibreC}

Before doing the computation of $\la \phi^2 \ra$ and $\la \phi z_L z_R \ra $, it will be helpful to present the computation of $\la \phi \ra$ from a different point of view. This will allow us to introduce the Ginibre ensemble of complex matrices \cite{Ginibre}, that we will use to compute $\la \phi^* \phi \ra$ and $\la \phi z_L z_R \ra $.

\vskip .3in

Let us consider the average $\la \phi(H_0) \ra_{H_0}$. In \eqref{phiAv} we showed that it is zero by first doing the $H_0$ integral and then $\delta \Sigma$ integral. In GUE, we can instead use \eqref{phiGUE2} and write
\begin{align}
\la \phi(H_0) \ra_{H_0} &= {L^2 \over \beta_L \beta_R} \oint {du_L \over 2\pi i} {du_R \over 2\pi i } ~ e^{-\beta_L u_R - \beta_R u_R} 
\nn
\\ 
&\times \left\la 
\left( e^{-\beta_L R_{H_0+i\delta\Sigma}(u_L)}  - e^{-\beta_L/u_L} \right)
\left( e^{  -\beta_R  R_{H_0 + i\delta\Sigma}(u_R)  }  - e^{-\beta_R/u_R} \right)
\right\ra_{H_0, \delta \Sigma} + \dots \ . 
\label{phiAv2}
\end{align}
Here, both $H_0$ and $\delta \Sigma$ are hermitian. The 2-matrix model of $H_0, \delta \Sigma$ effectively becomes a matrix model of the complex matrix $M = H_0 + i \delta \Sigma$. To compute \eqref{phiAv2} we are interested in correlators of $M$ of the type
\begin{align} \label{GinibreEns}
\int dH_0 d\delta \Sigma ~ e^{- {L\over 2} \tr (H_0^2 + \delta \Sigma^2)} f(H_0 + i \delta \Sigma) &= 
\int dM dM^\dagger ~ e^{- {L\over 2} \tr M^\dagger M} f(M)  \\
& = f(0) \ , 
\end{align}
where the measure is $dM dM^\dagger  = \prod_{i,j} d \re M_{ij} d \im M_{ij}$. We assumed the function $f(M)$ is holomorphic in $M$. The integration over the phases of matrix elements $M_{ij}$ picks out only the $M$-independent part $f(0)$. In particular, the correlators of resolvents trivially factorize
\begin{align}\label{GinRcorr1}
\left\la 
R_{H_0 + i \delta \Sigma}(u_1)  \dots R_{H_0 + i \delta \Sigma}(u_n) 
\right\ra_{H_0, \delta \Sigma}& = 
\left\la 
R_{H_0 + i \delta \Sigma}(u_1)
\right\ra_{H_0, \delta \Sigma}     
\dots 
\left\la 
R_{H_0 + i \delta \Sigma}(u_n) 
\right\ra_{H_0, \delta \Sigma}  \\ 
&= {1\over u_1} \dots {1\over u_n} \ .
\label{GinRcorr2}
\end{align}
Using this in \eqref{phiAv2}, we immediately find $\la \phi \ra = 0$, as expected. 

\vskip .3in

The gaussian ensemble of complex matrices \eqref{GinibreEns} was studied by Ginibre \cite{Ginibre} long time ago (also see chapter 15 in \cite{Mehta}). The equations \eqref{GinRcorr1}, \eqref{GinRcorr2} is the only property of this ensemble that we will use. Though we will not need this, similarly to GUE one can reduce the matrix integral over $M$ to an integral over eigenvalues only and study correlators of the density of complex eigenvalues \cite{Ginibre,Mehta}.\footnote{Note, however, that using a unitary transformations one cannot in general reduce a complex matrix to a diagonal form. Instead, the most one can do is to reduce it to an upper triangular form $M = U(\Lambda + T)U^\dagger$, where $\Lambda$ is a diagonal matrix of eigenvalues of $M$ and $T$ is strictly upper triangular. This is Schur decomposition. Holomorphic observables like $\tr M^n$ depend only on the eigenvalues and can be derived from correlators of the density of eigenvalues. On the other hand, non-holomorphic observables like $\tr M^\dagger M$ also depend on $T$ and cannot be computed from the density of eigenvalues. See e.g. sections 1.2 and 3.4.3 in \cite{Eynard:2015aea}. }

\subsection{Computation of $\la \phi^2 \ra$}

Now we would like to compute $\la \phi^2 \ra, \la \phi z_L z_R \ra$ and derive \eqref{phi2Res}. From \eqref{phiGUE2} we have
\begin{align}
 \label{AvPhi2MM0}
\la \phi(H_0)^2 \ra_{H_0} \approx& \left( L^2 \over \beta_L \beta_R \right)^2 \oint
{du_L du_R d  u_{L'} d  u_{R'} \over (2\pi i)^4 } ~ 
 e^{-\beta_L (u_L +  u_{L'})-\beta_R (u_R+ u_{R'} ) } \\
 & \times 
 \big\la 
 \left( e^{-\beta_L R_{H_0+i\delta\Sigma}(u_L)}  - e^{-\beta_L/u_L} \right)
\left( e^{  -\beta_R  R_{H_0 + i\delta\Sigma}(u_R)  }  - e^{-\beta_R/u_R} \right) \\ 
&\qquad
\left( e^{-\beta_L R_{H_0+i\delta\Sigma'}( u_{L'})}  - e^{-\beta_L/  u_{L'}} \right)
\left( e^{  -\beta_R  R_{H_0 + i\delta\Sigma'}(  u_{R'})  }  - e^{-\beta_R/ u_{R'}} \right)
 \big\ra_{H_0, \delta \Sigma, \delta \Sigma'} \ .
 \label{AvPhi2MM}
\end{align}
There are four factors corresponding to the four systems $L,R,L', R'$. We need to do the averaging over $H_0, \delta \Sigma, \delta \Sigma'$. Our strategy for computing this average is as follows. First, we compute the $H_0$ average, while keeping $\delta \Sigma, \delta \Sigma'$ fixed. Then we do the average over $\delta \Sigma, \delta \Sigma'$.

\vskip .3in

In the first step, we average over $H_0$, while keeping $\delta \Sigma, \delta \Sigma'$ fixed. We can think of this as considering an ensemble of hamiltonians given by a sum of the deterministic and random parts. Namely, in \eqref{AvPhi2MM} we have two distinct matrices: $H_1 = H_0 + i \delta \Sigma$ and $H_2 = H_0 + i \delta \Sigma'$. In the first step we think of $H_0$ as the random part, while $\delta \Sigma, \delta \Sigma'$ are deterministic parts, i.e. fixed. 

\vskip .3in

Correlation functions in such a ``deterministic plus random'' ensemble have been studied by Brezin, Hikami and Zee \cite{Brezin:1993lnq,Brezin:1994sq,Brezin:1995fs}. In particular, they computed the connected two-point function in GUE at large $L$
\begin{align}\label{RR det+rand}
\Big\la R_{H_0 + i \delta \Sigma}(u) R_{H_0 + i \delta \Sigma'}(u') \Big\ra_{H_0,c} &=
 - {1\over L^2} \p_u \p_{u'} \log\Big( 1 - {\cal F}_{\delta \Sigma, \delta \Sigma'}(u,u') \Big) \ , 
 \end{align}
 where 
 \begin{align}
 {\cal F}_{\delta \Sigma, \delta \Sigma'}(u,u')  &\equiv  {1\over L} 
\tr \left(  {1\over  u - i \delta \Sigma -  \la R_{H_0 + i \delta \Sigma} (u)\ra_{H_0}  }  ~ { 1 \over  u' - i \delta \Sigma' -  \la R_{H_0 + i \delta \Sigma'}(u') \ra_{H_0}  } \right)  \ .
\label{circDef}
\end{align}

A formula equivalent to \eqref{RR det+rand} was first derived in \cite{Brezin:1993lnq} by summing over planar diagrams. The method is essentially unchanged from the more standard case $\delta \Sigma = \delta \Sigma' = 0$. Later, in \cite{Brezin:1994sq}, it was noticed that the result can be written in a particularly simple form \eqref{RR det+rand}.\footnote{In \cite{Brezin:1993lnq,Brezin:1994sq} the authors derived \eqref{RR det+rand} for $H_0$ drawn from GUE. In \cite{Brezin:1995fs}, it was generalized to a matrix model with an arbitrary potential, albeit only when the deterministic parts are equal $\delta \Sigma =  \delta \Sigma'$. It is natural to expect that a similar result holds for distinct $\delta \Sigma, \delta \Sigma'$ and a general matrix potential. } One can check that for $\delta \Sigma = \delta \Sigma' = 0$ one recovers the usual two-point function of resolvents \cite{Ambjorn:1990ji, Brezin:1993qg}.

\vskip .3in

Equipped with the result \eqref{RR det+rand}, we now return to the computation of \eqref{AvPhi2MM}. This computation is conceptually similar to the analogous computation in \cite{Saad:2021rcu,Mukhametzhanov:2021nea}. There are four factors in \eqref{AvPhi2MM} representing the systems $L,R,L',R'$. The important contributions come from pairs of wormholes between these systems, $(LL')(RR')$ and $(LR')(RL')$ 
\begin{align}\label{HW}
\la \phi^2 \ra  \qquad = \qquad  \includegraphics[scale=.35, valign=c]{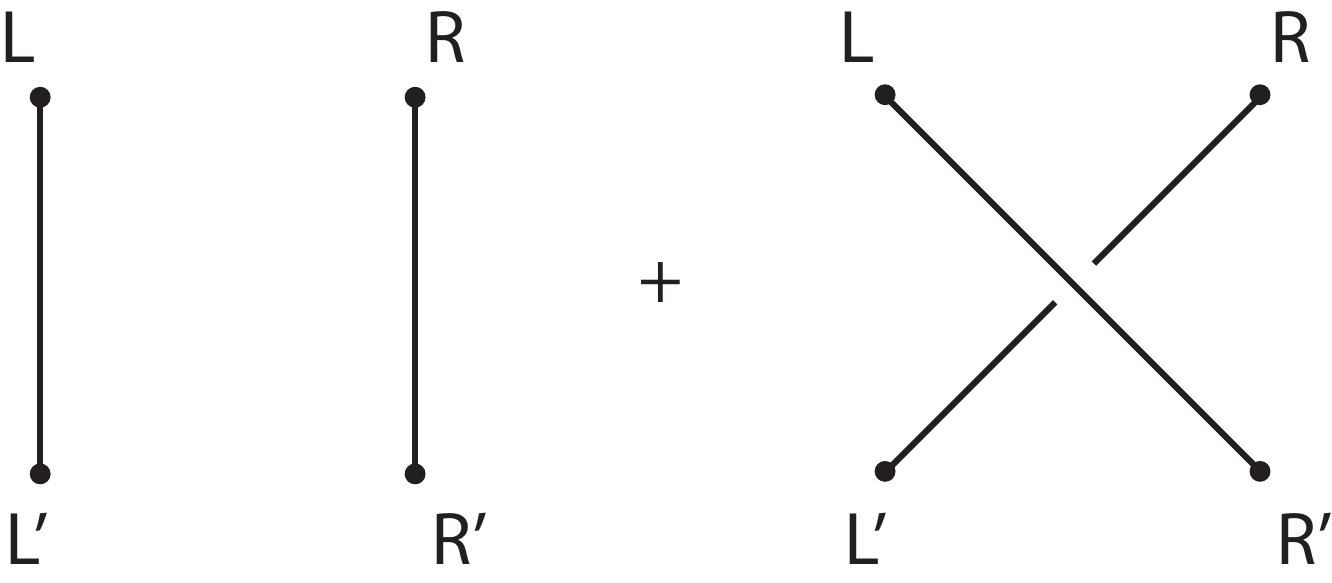} 
\end{align}
The correlation pattern $(LR)(L'R')$ is absent because it corresponds to the disconnected term $\la \phi \ra \la \phi \ra =0$.

\vskip .3in

We first do the average over $H_0$ and include terms corresponding to the two figures in \eqref{HW} 
\begin{align}
 &\big\la 
 \left( e^{-\beta_L R_{H_0+i\delta\Sigma}(u_L)}  - e^{-\beta_L/u_L} \right)
\left( e^{  -\beta_R  R_{H_0 + i\delta\Sigma}(u_R)  }  - e^{-\beta_R/u_R} \right) \\ 
&\qquad
\left( e^{-\beta_L R_{H_0+i\delta\Sigma'}( u_{L'})}  - e^{-\beta_L/  u_{L'}} \right)
\left( e^{  -\beta_R  R_{H_0 + i\delta\Sigma'}(  u_{R'})  }  - e^{-\beta_R/ u_{R'}} \right)
 \big\ra_{H_0}  \\
 \approx &
 \label{LLRRExp}
 \left\la e^{-\beta_L R_{H_0+i\delta\Sigma}(u_L)}  \right\ra_{H_0}
 \left\la e^{  -\beta_R  R_{H_0 + i\delta\Sigma}(u_R)  }  \right\ra_{H_0}
 \left\la e^{-\beta_L R_{H_0+i\delta\Sigma'}( u_{L'})}  \right\ra_{H_0}
 \left\la  e^{  -\beta_R  R_{H_0 + i\delta\Sigma'}(  u_{R'})  }  \right\ra_{H_0} \\ 
 & \qquad \times (\beta_L \beta_R)^2 
 \left\{
 \left\la R_{H_0 + i \delta \Sigma}(u_L) R_{H_0 + i \delta \Sigma'}( u_{L'}) \right\ra_{H_0,c} 
~ \left\la R_{H_0 + i \delta \Sigma}(u_R) R_{H_0 + i \delta \Sigma'}( u_{R'}) \right\ra_{H_0,c} 
+(L' \leftrightarrow R')
 \right\} + \dots  \ ,
 \label{LLRRH0}
\end{align}
where ellipsis contain terms that vanish after averaging over $\delta \Sigma, \delta \Sigma'$ and terms suppressed at large $L$. The two-point functions of resolvents in \eqref{LLRRH0} can be computed using \eqref{RR det+rand}. 

\vskip .3in

Now we take the averages over $\delta \Sigma, \delta \Sigma'$. Note that the two-point function \eqref{RR det+rand} is of order $1\over L^2$. Two such factors in \eqref{LLRRH0} cancel $L^4$ factor in \eqref{AvPhi2MM0} and the result is of $O(1)$. We are interested only in the leading $O(1)$ terms. Therefore, when we take the average over $\delta \Sigma, \delta \Sigma'$, it is sufficient to include only one-point functions of resolvents. Omitting various indices, $\la R \ra \sim 1$. While higher-point correlators of resolvents in $\delta \Sigma, \delta \Sigma'$ are suppressed at large $L$ by the usual genus counting.

\vskip .3in

With this understanding, averaging \eqref{LLRRH0} over $\delta \Sigma, \delta \Sigma'$ amounts to averaging each resolvent separately. First, there are exponentials \eqref{LLRRExp} that are simple to compute using \eqref{GinRcorr2}
\begin{align}\label{AvExp}
 \left\la e^{-\beta R_{H_0+i\delta\Sigma}(u)}  \right\ra_{H_0, \delta \Sigma} = e^{-\beta/u} 
 \end{align}
Second, we approximate
\begin{align}\label{AvRRLog}
\left\la  \left\la R_{H_0 + i \delta \Sigma}(u) R_{H_0 + i \delta \Sigma'}( u') \right\ra_{H_0,c} \right\ra_{\delta \Sigma, \delta \Sigma'} \approx -{1\over L^2}
\p_u \p_{u'} \log\left( 1 - \left\la {\cal F}_{\delta \Sigma, \delta \Sigma'}(u,u') \right\ra_{\delta \Sigma, \delta \Sigma'}   \right) \ .
\end{align}
Using the definition \eqref{circDef}, we compute
\begin{align}\label{Fav}
  \left\la {\cal F}_{\delta \Sigma, \delta \Sigma'}(u,u') \right\ra_{\delta \Sigma, \delta \Sigma'} 
  &=
 {1\over L} 
 \left\la 
 \tr {1\over u - i \delta \Sigma -  \la R_{H_0 + i \delta \Sigma} (u) \ra_{H_0} }
 ~
 {1 \over  u' - i \delta  \Sigma' -  \la R_{H_0 + i \delta \Sigma'} (u')\ra_{H_0}  } 
 \right\ra_{\delta \Sigma, \delta \Sigma'} \\
 &=
 {1\over L^2}  \left\la 
 \tr {1\over u - i \delta \Sigma -   \la R_{H_0 + i \delta \Sigma} (u) \ra_{H_0} } \right\ra_{\delta \Sigma}
 ~ 
\left\la 
 \tr {1\over u' - i \delta \Sigma' - \la R_{H_0 + i \delta \Sigma'} (u')\ra_{H_0} } 
 \right\ra_{\delta \Sigma'} \ .
\label{Fav2}
\end{align}
In the second line, we integrated over the relative unitary between $\delta \Sigma, \delta \Sigma'$. In more detail, we diagonalize by unitary transformations $\delta \Sigma = U^\dagger \delta \sigma U, \delta \Sigma' = V^\dagger \delta \sigma' V $. Note that $ \la R_{H_0 + i \delta \Sigma} (u) \ra_{H_0} $ depends only on the eigenvalues of $\delta \Sigma$. Then integrate in \eqref{Fav} over the relative unitary $W = U V^\dagger$ and use that $\int dW ~ W^\dagger_{ij} W_{kl} = {1\over L} \delta_{jk}\delta_{il}$.

\vskip .3in

To leading order, the function $ \la R_{H_0 + i \delta \Sigma} (u) \ra_{H_0}$ in the denominator of \eqref{Fav2} can be substituted by $ \la R_{H_0 + i \delta \Sigma} (u) \ra_{H_0, \delta \Sigma} = {1\over u}$. Thus, we find
\begin{align}
 \left\la {\cal F}_{\delta \Sigma, \delta \Sigma'}(u,u') \right\ra_{\delta \Sigma, \delta \Sigma'} 
\approx 
\left\la 
R_{i\delta \Sigma}\left( u - {1\over u} \right)
\right\ra_{\delta \Sigma}
\left\la 
R_{i\delta \Sigma'}\left( u' - {1\over u'} \right)
\right\ra_{\delta \Sigma'} \ .
\end{align}
The expression for the GUE resolvent is well-known $\la R_{M}(w) \ra_{M} = {1\over 2}(w - \sqrt{w^2 - 4})$. Therefore
\begin{align}\label{AvCirc}
\left\la {\cal F}_{\delta \Sigma, \delta \Sigma'}(u,u') \right\ra_{\delta \Sigma, \delta \Sigma'} 
\approx 
{1\over u u'} \ .
\end{align}
We made a particular choice of the branch of the square root corresponding to $|u| > 1, |u'| >1$. Different choices would lead to the same result, that differs only by a change of coordinates in the integral \eqref{AvPhi2MM}. 

\vskip .3in

Finally, using \eqref{AvCirc}, the two-point function of resolvents \eqref{AvRRLog} takes a simple form 
\begin{align}\label{AvRRLog2Phi2}
\left\la  \left\la R_{H_0 + i \delta \Sigma}(u) R_{H_0 + i \delta \Sigma'}( u') \right\ra_{H_0,c} \right\ra_{\delta \Sigma, \delta \Sigma'} \approx 
{1\over L^2} {1\over (1-u u')^2}
\end{align}
Combining equations \eqref{LLRRH0}, \eqref{AvExp}, \eqref{AvRRLog2Phi2} and inserting them in \eqref{AvPhi2MM}, we find
\begin{align}
\la \phi^2 \ra \approx 
\oint
{du_L du_R d  u_{L'} d  u_{R'} \over (2\pi i)^4 } ~ 
 &e^{
 -\beta_L \left(u_L + {1\over u_L} \right)
 -\beta_L \left(  u_{L'} +  {1\over  u_{L'}}  \right)
  -\beta_R \left(u_R + {1\over u_R} \right)
 -\beta_R \left(  u_{R'} +  {1\over  u_{R'}}  \right)
 } 
 \\
   \times &
\left\{ 
{1\over (u_L -  u_{L'})^2} {1\over (u_R -  u_{R'})^2}
+ 
{1\over (u_L -  u_{R'})^2} {1\over (u_R -  u_{L'})^2}
\right\} \ ,
\label{Phi2Final}
\end{align}
where we also changed variables for $L', R'$ by $ u_{L'} \to {1\over u_{L'}} \ ,   u_{R'} \to {1\over  u_{R'}}$.

\vskip .3in

We recognize the two terms in \eqref{Phi2Final} as pairs of ``wormholes'', i.e. the universal two-point correlators in a single-cut matrix model. Precisely these integrals were computed in \cite{Brezin:2007aa, Blommaert:2021gha} (see sections 2.2 and 3 respectively in those papers). Those references showed that integrals \eqref{Phi2Final} are equivalent to the Laplace transform of the universal density-density correlator \cite{Ambjorn:1990ji, Brezin:1993qg}
\begin{align}\label{utoE}
&\oint {du du' \over (2\pi i)^2} ~ e^{-\beta\left( u + {1\over u} \right) -\beta'\left( u' + {1\over u'} \right)} {1\over (u-u')^2} 
=\int_{-2}^2 dE dE' ~ e^{-\beta E - \beta' E'} \la \rho(E) \rho(E') \ra_c \ , \\
& \la \rho(E) \rho(E') \ra_c 
=
-{1\over 2\pi^2} {1\over (E-E')^2} { 4-E E'  \over \sqrt{(4-E^2)(4-E'^2)}} \ .
\label{den2Corr}
\end{align} 
We give an alternative derivation in appendix \ref{uoint}. We therefore derived one of the equations \eqref{phi2Res}
\begin{align}
\la \phi^2  \ra \approx 
\la z_L z_{L'} \ra \la z_R z_{R'} \ra +
 \la z_L z_{R'} \ra \la z_R z_{L'} \ra  \ .
\end{align}

\subsection{Computation of $\la \phi z_L z_R \ra$}

The computation of $\la \phi z_L z_R \ra$ is very similar to $\la \phi^2 \ra$. First, we write the partition functions $z_L, z_R$ as integrals of the resolvent
\begin{align}
Z(H_0) = \tr e^{-\beta H_0} = L \oint {dw \over 2\pi i } ~ e^{-\beta w} R_{H_0}(w) \ .
\end{align} 
where the contour goes around the eigenvalues of $H_0$. Using \eqref{phiGUE2}, we write
\begin{align}
\label{phiZZ}
\la \phi(H_0) z_L(H_0) z_R(H_0) \ra_{H_0} &\approx 
{ L^4 \over \beta_L \beta_R } \oint
{du_L du_R d  w_{L'} d  w_{R'} \over (2\pi i)^4 } ~ 
 e^{-\beta_L (u_L +  w_{L'})-\beta_R (u_R+ w_{R'} ) } \\
 & \times 
 \big\la 
 \left( e^{-\beta_L R_{H_0+i\delta\Sigma}(u_L)}  - e^{-\beta_L/u_L} \right)
\left( e^{  -\beta_R  R_{H_0 + i\delta\Sigma}(u_R)  }  - e^{-\beta_R/u_R} \right) \\
 & \qquad \left( R_{H_0}(w_{L'})  +{1\over \beta_L} e^{-\beta_L/w_{L'}} \right)
 \left( R_{H_0}(w_{R'})  +{1\over \beta_R}e^{-\beta_R/w_{R'}} \right)
 \big\ra_{H_0, \delta \Sigma} \ .
\end{align}
This is similar to \eqref{AvPhi2MM}, except that for systems $L',R'$ we have resolvents instead of exponentials of resolvents and $\delta \Sigma' = 0$.

Again, we first compute the $H_0$ average. Similarly to the previous subsection, the two important contributions come from the correlations $(LL')(RR')$ and $(LR')(RL')$
\begin{align}
&\big\la 
 \left( e^{-\beta_L R_{H_0+i\delta\Sigma}(u_L)}  - e^{-\beta_L/u_L} \right)
\left( e^{  -\beta_R  R_{H_0 + i\delta\Sigma}(u_R)  }  - e^{-\beta_R/u_R} \right) \\
 & \qquad \times \left( R_{H_0}(w_{L'})  +{1\over \beta_L} e^{-\beta_L/w_{L'}} \right)
 \left( R_{H_0}(w_{R'})  +{1\over \beta_R} e^{-\beta_R/w_{R'}} \right)
 \big\ra_{H_0} \\ 
  \approx &
  \left\la e^{-\beta_L R_{H_0+i\delta\Sigma}(u_L)}  \right\ra_{H_0}
 \left\la e^{  -\beta_R  R_{H_0 + i\delta\Sigma}(u_R)  }  \right\ra_{H_0} \\
& \qquad   \times \beta_L \beta_R 
 \left(
 \left\la R_{H_0 + i \delta \Sigma}(u_L) R_{H_0}( w_{L'}) \right\ra_{H_0,  c} 
~ \left\la R_{H_0 + i \delta \Sigma}(u_R) R_{H_0 }( w_{R'}) \right\ra_{H_0,c} 
+(L' \leftrightarrow R')
 \right) + \dots \ .
\end{align}
Now we compute the $\delta \Sigma$ average similarly to \eqref{AvRRLog}
\begin{align}\label{AvRRLog2}
\left\la  \left\la R_{H_0 + i \delta \Sigma}(u) R_{H_0 }( w) \right\ra_{H_0,c} \right\ra_{\delta \Sigma} \approx -{1\over L^2}
\p_u \p_{w} \log\left( 1 -  \left\la {\cal F}_{\delta \Sigma, \delta \Sigma'=0}(u,w) \right\ra_{\delta \Sigma}  \right) \ ,
\end{align}
where ${\cal F}_{\delta \Sigma, \delta \Sigma'}$ is defined in \eqref{circDef}. We find
\begin{align}
  \left\la {\cal F}_{\delta \Sigma, \delta \Sigma'=0}(u,w) \right\ra_{\delta \Sigma} 
 & =
 {1\over L} 
 \left\la 
 \tr
 {1\over u - i \delta \Sigma - \la R_{H_0+ i \delta \Sigma}(u) \ra_{H_0} }
 \right\ra_{\delta \Sigma}
~ {1 \over  w - \la R_{H_0}(w) \ra_{H_0}   } 
\\
&\approx 
\Big\la 
R_{i\delta \Sigma}\left( u - {1\over u} \right)
\Big\ra_{\delta \Sigma} ~ {1 \over  w - \la R_{H_0}(w) \ra_{H_0}   }  \\
&\approx {1\over u} ~ {w - \sqrt{w^2 - 4} \over 2} \\
& = {1\over u u'} \ , \qquad \left( w = u' + {1\over u'}, ~ |u'| > 1 \right)
\end{align}
In the second line, we approximated $\la R_{H_0+ i \delta \Sigma}(u) \ra_{H_0}$ by the average $\la R_{H_0+ i \delta \Sigma}(u) \ra_{H_0, \delta \Sigma} = {1\over u}$. In the third line, we used the GUE resolvent $\la R_M(w) \ra_M = {1\over 2}(w- \sqrt{w^2 - 4})$ and assumed $|u| > 1$. In the fourth line, we assumed $|u'| >1$ to invert the Zhukovsky transformation.

Inserting these results into \eqref{phiZZ} and changing variables to $u_{L'}, u_{R'}$ by $w_a = u_a + {1\over u_a}, ~  a = (L', R') $, we find an expression identical to \eqref{Phi2Final}. Therefore, we derived the second equation \eqref{phi2Res}
\begin{align}
\la \phi z_L z_R \ra \approx 
\la z_L z_{L'} \ra \la z_R z_{R'} \ra +
 \la z_L z_{R'} \ra \la z_R z_{L'} \ra  \ .
\end{align}

We now derived both equations \eqref{phi2Res} and showed that the errors in the approximation \eqref{HWapproxMM2} are indeed small \eqref{ErrorSmall} in GUE for a typical choice of the hamiltonian.

\section{Discussion}
\label{discuss}

We considered the factorization problem in SYK and in Matrix Models, viewed as toy models of holographic quantum gravity. In a theory with fixed couplings $j$, we introduced a fictitious ensemble averaging over couplings $J$ by inserting a delta function $\delta(J-j)$. For a typical choice of the couplings $j$, we isolated two leading contributions: ``wormhole'' and the ``pair of linked half-wormholes''. 

\vskip .3in

In the SYK model this gave a different derivation of the result in \cite{Saad:2021rcu}. Our method, however, allows a straightforward generalization to other models with an ensemble average. In particular, we considered a matrix model with an arbitrary potential and proposed an expression for the pair of linked half-wormholes \eqref{phiDef}.

\vskip .3in

In GUE, we were able to explicitly compute the contribution of the pair of linked half-wormholes and show that together with the wormhole it gives a good approximation in the theory with fixed couplings.

\vskip .3in

Finally, in SYK and in GUE we noted a curious interpretation that the pair of linked half-wormholes can be thought of as the squared partition function averaged over the imaginary part of the couplings, see \eqref{ImJ}, \eqref{phiSYKim}, \eqref{phiGUEim}.

\subsection{Large $N$ factorization}

To show that the approximation including the wormhole and the pair of linked half-wormholes is valid in a theory with fixed couplings, we needed to compute the averages $\la \phi^2 \ra, \la \phi z_L z_R \ra$. In both SYK with one time point \cite{Saad:2021rcu,Mukhametzhanov:2021nea} and in GUE it turned out that to leading order
\begin{align}\label{LargeNFact}
\la \phi^2 \ra  \approx \la \phi z_L z_R \ra \approx \la z_L z_R \ra^2 + \la z_L^2 \ra \la z_R^2 \ra \ ,
\end{align}
where $\phi$ is the pair of linked half-wormholes. This result looks quite general and one might expect it is true in other large $N$ theories with an ensemble average. Moreover, we showed that in some cases $\phi$ can be thought of as the squared partition function averaged over the imaginary part of the couplings. From this point of view, the property \eqref{LargeNFact} looks like some sort of large $N$ factorization. In particular, this makes us believe \eqref{LargeNFact} might hold in both SYK and in matrix models with arbitrary potentials. In turn, this would justify the steps outlined in \eqref{zzDeltaIntro}, \eqref{zzWIntro} and resolve the factorization problem in these cases.

\subsection{Fixed couplings vs. dynamical gravity}

There seems to be an apparent conflict between a theory with fixed couplings and dynamical gravity in the bulk. On the one hand, there is no doubt that AdS/CFT can be true in theories without ensemble averages. This is certainly the case in examples in higher dimensions, such as $N=4$ SYM or ABJM \cite{Maldacena:1997re, Aharony:2008ug}. On the other hand, in lower dimensional examples such as JT gravity \cite{Saad:2019lba} or minimal string theory \cite{David:1984tx, Ambjorn:1985az, Kazakov:1985ds, Kazakov:1985ea, Brezin:1990rb, Douglas:1989ve, Gross:1989vs, Seiberg:2003nm, Maldacena:2004sn}, the ensemble averaging seems to be intimately tied with dynamical gravity in the bulk. Namely, one can think of fluctuating surfaces in the bulk as arising from triangulations by double-line `t Hooft graphs in random matrix perturbation theory (see \cite{Ginsparg:1993is, DiFrancesco:1993cyw} for a review).\footnote{In SYK, the emergence of dynamical gravity in the bulk can be interpreted in a similar way. In this case the hamiltonian is only partially random. Out of $2^N$ matrix elements of the hamiltonian roughly $N^4$ are random, while others are fixed. We can think of this partially random hamiltonian as leading to fluctuating surfaces in the bulk through the triangulation picture.} So naively, at least in these low-dimensional examples, it seems that if we fix the couplings we loose dynamical gravity in the bulk. A similar connection between ensemble averages of 2d CFTs and dynamical gravity in 3d was recently explored in \cite{Maloney:2020nni, Afkhami-Jeddi:2020ezh, Cotler:2020ugk, Dong:2021wot, Collier:2021rsn}.

\vskip .3in 

A possible resolution of this apparent conflict is a formula like \eqref{Z2int}, which we repeat here
\begin{align}\label{2-MatrixModel}
Z_L(H_0) Z_R (H_0)
 = e^{L \tr V(H_0) } \int dH d\Sigma ~e^{- L \tr   V(H)  }   e^{iL \tr \Sigma(H-H_0)}~Z_L(H) Z_R (H) \ .
\end{align}
On the one hand, the LHS has a manifest description as a theory with a fixed hamiltonian (couplings) $H_0$. On the other hand, the description on the RHS has fluctuating couplings $H$ dual to a putative dynamical gravity in the bulk, as well as extra degrees of freedom dual to $\Sigma$.

\subsection{Open questions}

There are several open questions. 

\vskip .3in

$\bullet$ The most immediate task is to explore matrix models with polynomial potentials $V(E)$, compute the contribution of the linked half-wormholes \eqref{phiDef} more explicitly and show that errors in the approximation \eqref{HWapproxMM2} are small. This requires computing random matrix correlators in the external field $\Sigma$. This question have been explored for a general matrix potential e.g. in \cite{Gross:1991aj, Zinn-Justin:1997hrj, 1998CMaPh.194..631Z}.

\vskip .3in

$\bullet$ It would be interesting to know what (if any) is the gravity dual of the 2-matrix model \eqref{2-MatrixModel}. This 2-matrix model looks similar to the 2-matrix model dual to $(3,p)$ minimal string theory \cite{Douglas:1989dd}. However, there are a few important differences. First, the coupling between the two matrices is purely imaginary in our case $i \tr \Sigma H$, while it is real in \cite{Douglas:1989dd}. Another feature of \eqref{2-MatrixModel} is the coupling to the external field $i \tr \Sigma H_0$. On the other hand, a simplifying property of \eqref{2-MatrixModel} in comparison to \cite{Douglas:1989dd} is that the potential for the matrix $\Sigma$ is absent $V(\Sigma) = 0$. It would be interesting to know if one could take a sensible double-scaling limit of \eqref{2-MatrixModel} and find a gravity dual.

\vskip .3in

$\bullet$ Our results could be generalized to other theories with an ensemble average. For example, one might consider ensemble averages over Narain moduli spaces \cite{Maloney:2020nni, Afkhami-Jeddi:2020ezh, Dong:2021wot, Collier:2021rsn} at large central charge $c$ and ask whether there is a description of factorization in a particular CFT in the moduli space that is similar to ours. It should be possible to repeat the steps outlined in \eqref{zzDeltaIntro}, \eqref{zzWIntro} and find an expression for the pair of linked half-wormholes.\footnote{We thank Scott Collier for a discussion on this.}

\vskip .3in

$\bullet$ An intriguing possibility is that half-wormholes are related to the black hole interior and the black hole singularity.\footnote{This was suggested by Yiming Chen and Juan Maldacena.} See the discussion in section 6.2 of \cite{Saad:2021uzi}.  We observed that the linked half-wormholes can be thought of as averaging over the imaginary part of the couplings \eqref{phiSYKim}, \eqref{phiGUEim}.\footnote{In matrix models, we showed this only in GUE. It is not clear whether this statement generalizes to matrix models with arbitrary potentials in this precise form, but we still expect that complexified couplings play a role in this more general case as well.} It is tempting to think that these imaginary couplings are related to the analytic continuation into the interior of the black hole.

\section*{Acknowledgements}

We are grateful to Juan Maldacena, Sridip Pal and especially Phil Saad for stimulating conversations. This work was supported by a grant from the Simons Foundation (651444, BM).

{\appendix

\section{Factorization of $\tr H_0^2\tr H_0^2$ in GUE}
\label{trH2}

In this appendix we discuss the factorization problem for $\tr H_0^2\tr H_0^2$ in GUE. An approximation similar to \eqref{HWapproxMM2} is
\begin{align}\label{trH2approx}
\left( \tr H_0^2 - L \right) \left( \tr H_0^2 - L \right) &\approx \la \tr H^2\tr H^2 \ra_c + \int d\Sigma ~ e^{-{L\over 2} \tr (\Sigma + i H_0)^2} ~ \left( {\big\la \left( \tr H^2 - L \right) W(H) \big\ra \over \la W \ra } \right)^2 \ , 
\end{align}
where we used that $\la \tr H^2 \ra = L$ and $\la W\ra = e^{-{L\over 2} \tr \Sigma^2}$. With the help of the propagator
\begin{align}
\la H_{ij} H_{kl} \ra = {1\over L} \delta_{il}\delta{jk}
\end{align}
we compute ($W(H) =  e^{i L \tr \Sigma H}$)
\begin{align}
 \la \tr H^2\tr H^2 \ra_c = 2 \ , \qquad
{\big\la \left( \tr H^2 - L \right) W \big\ra  \over \la W \ra} = \tr (i \Sigma)^2 \ .
\end{align}
Inserting this into \eqref{trH2approx} and integrating over $\Sigma$ we have
\begin{align}\label{trH2approx2}
\left( \tr H_0^2 - L \right) \left( \tr H_0^2 - L \right) &\approx 
2 + \left\{ \left( \tr H_0^2 - L \right)^2 - 2 - 4\left( {1\over L} \tr H_0^2 - 1 \right)   \right\} \ .
\end{align}
Here, the first term ``$2$'' in the RHS is the ``wormhole'' contribution and the rest is the ``pair of linked half-wormholes''. The difference between the LHS and RHS in \eqref{trH2approx2} is 
\begin{align}
\text{Error}(H_0) =  - 4\left( {1\over L} \tr H_0^2 - 1 \right)  \ . 
\end{align}
It is straightforward to show that $\text{Error}$ is indeed small for a typical choice of $H_0$ drawn from a gaussian ensemble
\begin{align}
\la \text{Error}(H_0) \ra_{H_0} = 0 \ , \qquad 
\la \text{Error}(H_0)^2 \ra_{H_0} = {32 \over L^2} \ll \left( \la  \tr H_0^2 \tr H_0^2 \ra_{H_0,c} \right)^2 = 4 \ .
\end{align}
We also checked that a similar result holds for $\tr H_0^3 \tr H_0^3$.

\section{Contour integral representation of $\la z_L z_R \ra$ }
\label{uoint}

In this section we derive \eqref{utoE} 
\begin{align}\label{uInt}
\oint {du_1 du_2 \over (2\pi i )^2 } ~ {1\over (u_1 - u_2)^2} ~ e^{-\beta_L\left( u_1 + {1\over u_1} \right) -\beta_R\left( u_2 + {1\over u_2} \right)} \approx \la z_L z_R \ra = \la Z_L Z_R \ra_c \ ,
\end{align}
where the integration contours are around $u_1 = u_2 = 0$. The RHS is computed from the connected density-density correlator \cite{Ambjorn:1990ji, Brezin:1993qg} in GUE to leading order at large $L$.

\vskip .3in

We will use that a partition function can be written as a contour integral of the resolvent
\begin{align}\label{ZoInt}
Z = \tr e^{-\beta H} = L \oint {dw \over 2\pi i } ~ e^{-\beta w} R(w) \ , \qquad R(w) = {1\over L}\tr {1\over w- H} \ ,
\end{align}
where the contour goes around all eigenvalues of $H$. The idea of the computation is to change coordinates in the integral \eqref{uInt} to bring the $\beta_L, \beta_R$ dependence to the form similar to \eqref{ZoInt} and then identify the correlator $\la R(w) R(w') \ra_c$ as the universal two-point function of resolvents in single-cut matrix models \cite{Brezin:1993qg}.

\vskip .3in

We make a change of variables in \eqref{uInt}
\begin{align}\label{Zhuk}
w_i = u_i + {1\over u_i}, \qquad i = 1,2\ .
\end{align}
The Zhukovsky transformation $w=u + {1\over u}$ maps the exterior (or the interior) of the unit circle $|u| > 1$ to $w \in {\mathbb C}\backslash [-2,2]$. The unit circle $|u| = 1$ is mapped to the real interval $[-2,2]$. 

\vskip .3in

First, deform the contours in \eqref{uInt} to $|u_i| >1$. In this region we can invert \eqref{Zhuk}
\begin{align}
u_i(w_i) = {1\over 2} (w_i + \sqrt{w_i^2 - 4}) \ , i = 1,2 \ ,
\end{align}
where the branch of the square root is chosen by requiring it has a positive imaginary part above the cut. Using  \eqref{ZoInt} in the RHS and an identity
\begin{align}
\p_{w_1} \p_{w_2} \log\big( u_1(w_1) - u_2(w_2) \big) = {1\over \big( u_1(w_1) - u_2(w_2) \big) ^2 } ~ {du_1 \over dw_1} {du_2 \over dw_2} 
\end{align}
in the LHS, the equation \eqref{uInt} becomes
\begin{align}
\label{wIntRR}
&\oint {dw_1 dw_2 \over (2\pi i )^2}  ~ e^{-\beta_L w_1 - \beta_R w_2} 
~ \p_{w_1} \p_{w_2} \log \left( {u_1(w_1) - u_2(w_2) \over w_1 - w_2}  \right) \\
=& L^2 \oint {dw_1 dw_2 \over (2\pi i )^2} ~ e^{-\beta_L w_1 - \beta_R w_2}  ~\la R(w_1) R(w_2) \ra_c 
\ ,
\end{align}
where in the LHS we added a term $\p_{w_1} \p_{w_2} \log (w_1 - w_2) =  {1\over (w_1 - w_2)^2}$ that integrates to zero. We identify the connected correlator of resolvents
\begin{align}
L^2 \la R(w_1) R(w_2) \ra_c &=  \p_{w_1} \p_{w_2} \log \left( {u_1(w_1) - u_2(w_2) \over w_1 - w_2}  \right) \\ 
& = {1\over 2(w_1 -w_2)^2} 
\left(
{w_1 w_2 - 4 \over \sqrt{(w_1^2 - 4)(w_2^2 - 4)}} - 1
\right) \ ,
\end{align}
which is indeed the correct result for a single-cut matrix model \cite{Ambjorn:1990ji, Brezin:1993qg}. Further, deforming the contours in \eqref{wIntRR} to the cut, we pick up discontinuities across the cut and obtain the universal density-density correlator \eqref{den2Corr}.

}

\bibliography{refs}

\end{document}